\documentclass[aps, rmp,  twocolumn,preprintnumbers,amsmath,amssymb,floatfix]{revtex4}

\usepackage{graphicx}% Include figure files
\usepackage{epstopdf}
\usepackage{dcolumn}% Align table columns on decimal point
\usepackage{bm}% bold math

\begin{document}

\title{Can entropy save bacteria?}

\author{Suckjoon Jun}
\email{sjun@cgr.harvard.edu}
\affiliation{FAS Center for Systems Biology, Harvard University, 7 Divinity Avenue, Cambridge, MA 02138, USA}

\begin{abstract}
This article presents a physical biology approach to understanding organization and segregation of bacterial chromosomes. The author uses  a ``piston'' analogy for bacterial chromosomes in a cell, which leads to a phase diagram for the organization of two athermal chains confined in a {\it closed} geometry characterized by two length scales (length and width). When applied to rod-shaped bacteria such as \emph{Escherichia coli}, this phase diagram predicts that, despite strong confinement, duplicated chromosomes will demix, \i.e., there exists a primordial physical driving force for chromosome segregation. The author discusses segregation of duplicating chromosomes using the concentric-shell model, which predicts that newly synthesized DNA will be found in the periphery of the chromosome during replication. In contrast to chromosomes, these results suggest that most plasmids will be randomly distributed inside the cell because of their small sizes. An active partitioning system is therefore required for accurate segregation of low-copy number plasmids. Implications of these results are also sketched, e.g., on the role of proteins, segregation mechanisms for bacteria of diverse shapes, cell cycle of an artificial cell, and evolution.
\end{abstract}

\date{\today}

\maketitle

\tableofcontents

\section{Introduction}
\label{introduction}

\begin{figure}[t]
 \centering
 \includegraphics[width=8.4cm]{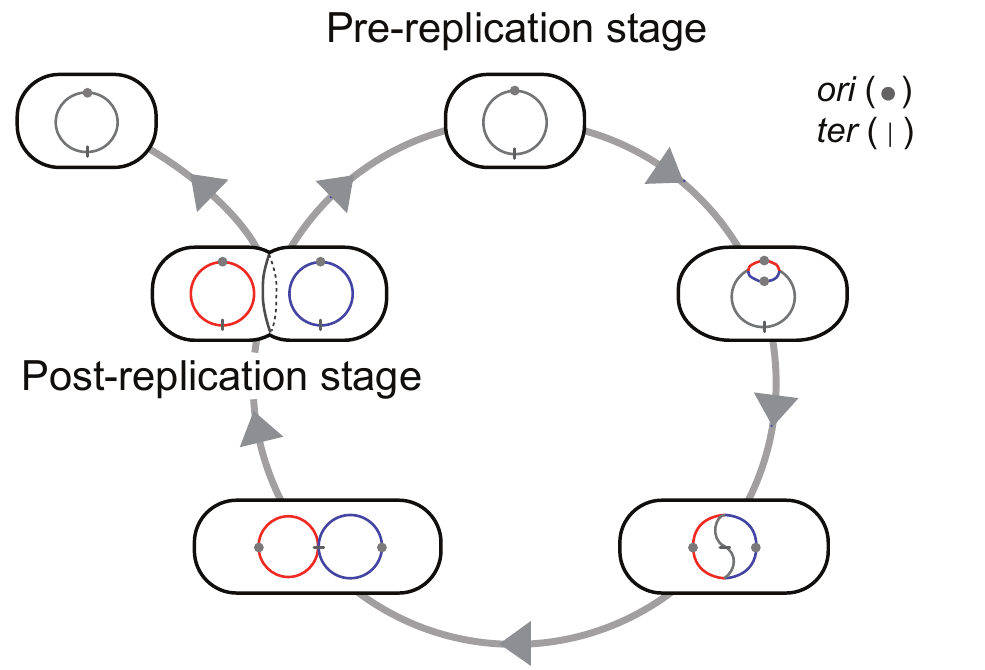}
 \caption{Schematic illustration of the bacterial cell cycle.  Replication begins at the 12' position (\emph{ori}) on the circular chromosome. Replication forks grow bidirectionally and meet at the opposite clock position (\emph{ter}). During the bacterial cell cycle, replication and segregation progress hand-in-hand, coupled with the growth of the cell.}
 \label{fig:cellcycle}
\end{figure}

Chromosomes are a cornerstone of fundamental processes of any cell, and harmony between their physical properties and biological functions is an evolutionary consequence. In bacteria, two defining processes of the cell cycle, DNA replication and chromosome segregation, progress hand-in-hand. The model system \emph{Escherichia coli} contains a single circular chromosome in a rod-shaped cell. DNA replication starts at a unique origin of replication (\emph{ori}), creating a replication bubble that grows bidirectionally, and the two replication forks meet at the terminus (\emph{ter}) located at approximately the opposite clock position of \emph{ori} on the circular chromosome (Fig.~\ref{fig:cellcycle}). In slowly growing cells, there is a one-to-one correspondence between each complete round of replication and the cell cycle. In fast growing cells, on the other hand, the cell divides more frequently than the progression of the forks from \emph{ori} to \emph{ter} and, thus, new replication has to initiate before the completion of the previous round of duplication, leading to ``multifork'' replication. In either case, replication and segregation are coordinated with the growth of the cell, and recent visualization experiments have begun to reveal how replicating chromosomes move and segregate during the bacterial cell cycle. Although details vary from organism to organism, the observation common to all bacteria of rod-shaped cells studied so far shows directed movement of duplicating chromosomes along the long axis of the cell [as seen in \emph{E. coli}~\citep{BatesKleckner05, Woldringh05, Wang06, Nielsen06}, \emph{Caulobacter crescentus}~\citep{Viollier04}, \emph{Bacillus subtilis}~\citep{Berkmen06} and \emph{Vibrio cholerae}~\citep{Fogel06, Fiebig06}]. Experimental data also suggest that, once duplicated, there is a fair degree of correlation between the relative clock position of circular chromosome and their \emph{average} positions along the long axis of the cell, which often (but not always) shows the principal linear ordering of chromosome~\citep{Wang06, Viollier04, Losick98}.

Perhaps the most influential model on chromosome segregation so far was proposed by ~\citet*{Jacob63} in their seminal paper on the replicon model of \emph{E. coli}: if replicating chromosomes (e.g., duplicated \emph{ori}'s) are attached to the elongating cell-wall membrane, they can be segregated passively by insertion of membrane material between the attachment points during growth. The model is intuitive and elegant -- but wrong. (See Sec.~\ref{subsec:shell}).

Since the work of Jacob and colleagues, two classes of models have been proposed to explain the observed directed drift of duplicating chromosomes: (a) biological mechanisms such as DNA replication~\citep{Grossman01}, co-transcriptional translation of membrane proteins~\citep{Woldringh02}, RNA transcription~\citep{Losick02}; and (b) physical/mechanical driving forces such as mechanical pushing between chromosomes~\citep{Kleckner04, BatesKleckner05} or conformational entropy of duplicating chromosomes~\citep{Danchin00, Jun06, Fan07}. These two classes of models are not mutually exclusive.

The second class of model (b) is important in that it provides a basic physical framework to examine the necessity and the roles of proteins or machinery (if any) involved in the cellular process and, if so, what conditions these putative biological players need to satisfy in order to successfully carry out their biological functions. For instance, when the cell is lysed, the bacterial chromosome expands to several times the size of the encapsulating cell~\citep{Woldringh01} (Fig.~\ref{fig:piston}A). This observation is puzzling since duplicated chromosomes must stay segregated without mixing until cell division for the viability of the cell, and yet the natural size of each chromosome is larger than the confining volume of the cell itself; why do the chromosomes not mix when they should occupy as much volume as possible? Thus, if the basic physical principles predict that the duplicated chromosomes should mix inside bacterial cell, this would explain why putative proteins must be actively involved in segregation of replicating chromosomes. However, if the physical analysis predicts otherwise, this would imply the existence of a primordial physical driving force underlying chromosome segregation in bacteria and that active biological mechanisms may be present for other reasons (e.g., when the cell cycle is perturbed by external stress).

The purpose of this paper is to present a comprehensive and rigorous physical model of bacterial chromosomes that makes experimentally testable predictions based on measurable physical parameters of the bacterial cell and its chromosomes.\footnote{For physical approach to eukaryotic chromosomes, the reader is encouraged read~\citet{MarkoSiggia97}.} We have developed our model with the idea that physical properties of the chromosomes must interplay with the shape and size of the cell, and bacteria might have evolved to learn how to regulate these parameters to ensure proper segregation of chromosomes. 
%As we shall show below, we have tested our model against the published data of \emph{E. coli}, and have used it to explain physically why duplicated chromosomes will stay demixed. A corollary of our result is that, because of their small sizes, plasmids will be randomly distributed within the cell unless they have a specific positioning mechanism, and we will discuss its implications.\\

This article is organized as follows: First, we begin with the concept of entropy, providing specific examples to explain how order can emerge from disorder driven by entropy. Second, we explain, using a piston analogy of bacterial chromosomes confined in a cell, how the geometric conditions of the cell interplay with the physical state of the confined chains. Based on this analogy, we present a phase diagram for the organization of self-avoiding chains in a box whose dimensions are defined by two lengthscales (width and length). To our knowledge, this is the first direct application of the de Gennes-Pincus blob theory~\citep{deGennesBook, Pincus76} to closed geometries with more than one length scale (as opposed to, e.g., open channels, slits, or spheres, which can be characterized by a single length scale).\footnote{In his celebrated book, {\it Scaling Concepts in Polymer Physics}, de Gennes noted ``This analysis can be extended to chains that are squeezed in slits and to other geometries provided that the confining object is characterized by a single length $D$.'' [see p. 51 in~\citet{deGennesBook}].} Then, we critically examine the model system \emph{E. coli} and predict that duplicated chromosomes of this organism should segregate spontaneously (with the ordering of individual chromosomes being principally linear unless interfered by tethering to the wall). We also predict the opposite behavior for plasmids (namely, random distribution) because of their small sizes. We discuss various implications concerning entropy-driven organization and segregation of chromosomes in bacteria of diverse shapes as well as in an artificial cell.\\

\section{Physical biology approach to bacterial chromosomes in a cell.}
\label{results}

\subsection{Entropy measures the degrees of freedom of the system}
\label{subsec:entropy}

In his influential book, {\it What is life?}, \citet{Schroedinger44} asserted that the only thermodynamically equilibrium state of a living being is death. Defining entropy as a measure of `disorder', in the last chapter `Is Life Based on the Laws of Physics?', he struggled to explain how order (life) can be achieved from disorder.  This association of life with minimal entropy, however, can be misleading.
\begin{figure}[tb]
 \centering
 \includegraphics[width=8.4cm]{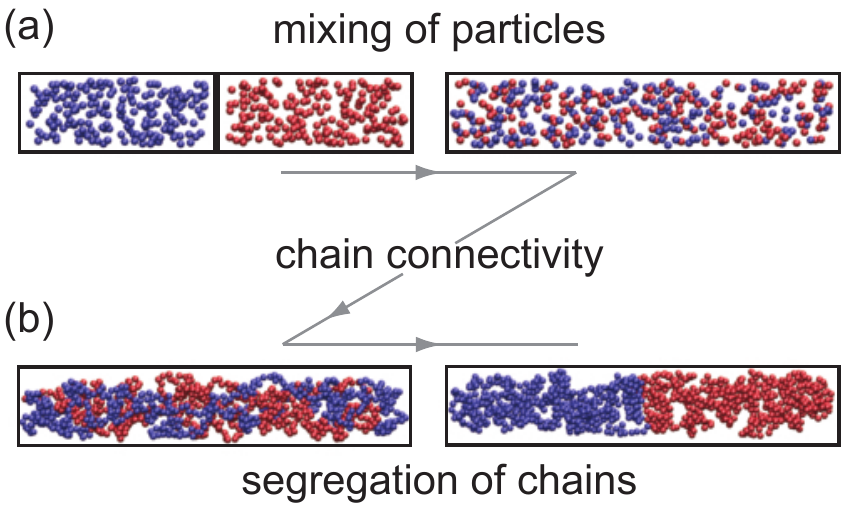}
 \caption{Subtle aspect of entropy. (Molecular dynamics simulation.) (a) Two species of particles ($N_1 = N_2 = 200$), initially separated by a wall, mix as the wall is removed. (b) If chain connectivity is introduced to this system by linearly connecting the particles of the same species, the two chains segregate. ($N_1 = N_2 = 512$)}
 \label{fig:entropy}
\end{figure}

In Fig.~\ref{fig:entropy}(a), we show snapshots from molecular dynamics simulations, where two species of $N_1$ and $N_2$ particles in equilibrium (blue and red, respectively) in a long rectangular box, initially separated by a wall. As we remove the wall from the system, the two species start to mix. The driving force of this process is the well-known ``entropy of mixing," which can be estimated as
\begin{eqnarray}
\Delta S_\mathrm{A} &=& k_B (\ln \Omega_\mathrm{A}^\prime - \ln \Omega_{\mathrm{A},0}) \\
	&\approx& -k_B \bigg[N_1 \ln\bigg(\frac{N_1}{N}\bigg) + N_2 \ln \bigg( \frac{N_2}{N} \bigg) \bigg] > 0,
\label{eq:mixing}
\end{eqnarray}
where $k_B$ is the Boltzman constant, $N = N_1 + N_2$, and $\Omega_\mathrm{A}^\prime = N!/N_1! N_2!$ ($\Omega_{\mathrm{A}, 0} = 1$) denotes the total number of configurations of the system after (before) mixing.

Entropy, however, is more subtle than a simple measure of disorder. To see this, let us start with the mixed state of Fig.~\ref{fig:entropy}(a) and connect the particles of the same species, creating two long linear chains, one painted with blue and the other with red [Fig.~\ref{fig:entropy}(b), left]. Another important condition is the excluded-volume interaction between the particles. The reader is encouraged to perform this simple computer simulation, and she or he will see that the two chains de-mix, \i.e., that ``order" emerges out of disorder (for the exact condition of segregation, see Sec.~\ref{subsec:phasediagram}).

There are also other examples in soft condensed matter physics where entropy leads to ordering. For instance, Onsager's hard-rod model~(\citeyear{Onsager49}) of the nematic-isotropic transition of liquid crystals is based on a similar physical insight of the trade-off between loss of orientational entropy and gain in positional entropy when hard rods are oriented parallel to one another. Crystallization of hard spheres \citep{Wood57, Alder57} is another example where ordering allows a larger room for fluctuations. 

The above examples clarify what entropy really measures, namely, the degrees of freedom, or the size of the phase space, of the system. For the main theme of this article, the specific case in Fig.~\ref{fig:entropy} points out the importance of chain connectivity, a keyword in polymer physics. In this example [Fig.~\ref{fig:entropy}(b)], chain connectivity defines a conformational space $\Omega_\mathrm{B}$ within the configuration space $\Omega_\mathrm{A}$, and within $\Omega_\mathrm{B}$ the typical conformations of the chains are those de-mixed. This emergence of order from disorder due to entropy is our starting point of understanding chromosome segregation in bacteria.\footnote{Note that polymer physics was still in its infant stage during Schr{\" o}dinger's time. For instance, Flory's magnum opus, {\it Principles of Polymer Chemistry} (\citeyear{Flory53}), appeared almost 10 years after {\it What is Life?} (\citeyear{Schroedinger44}).}

\subsection{Piston analogy of bacterial cells and chromosomes.}
\label{subsec:piston}

\begin{figure}[t]
 \centering
 \includegraphics[width=8.6cm]{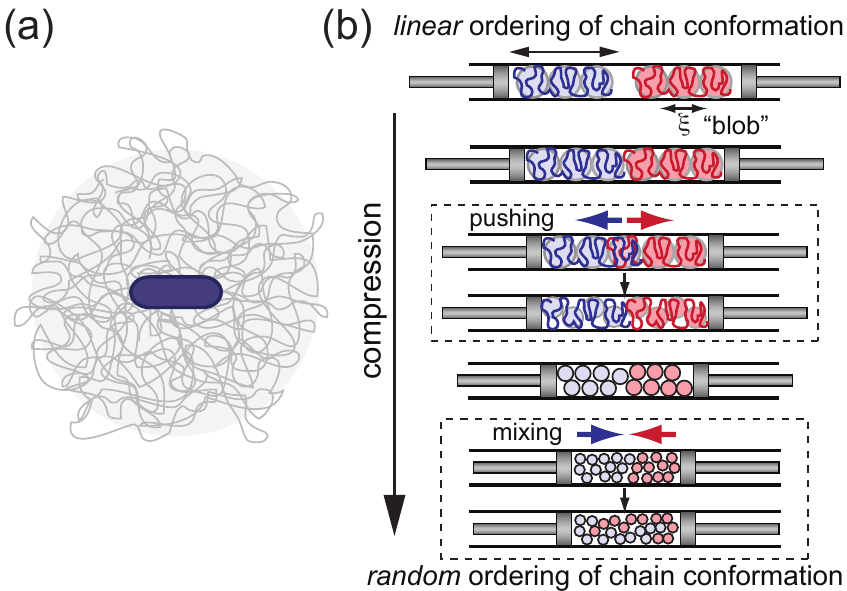}
 \caption{Piston analogy of bacterial chromosomes in confinement. (a) When the cell is lysed, the isolated chromosomes expands to several times the size of the cell. For our analysis, we used the published values of the nucleoid size before and after lysis, 0.24 $\mu$m~\citep{WoldringhOdijk} and 3.3$\mu$m~\citep{Oleg07}, respectively, for \emph{E. coli} B/r strain. (b) Two athermal chains (\i.e., with excluded-volume in good solvent) are confined in a cylinder whose pore width much smaller than the size of an unperturbed chain. The organization of the chain depends on the aspect ratio of the cylinder and the density of the monomers.}
 \label{fig:piston}
\end{figure}

Imagine a cylinder containing two long linear molecules with excluded-volume, closed by two pistons which are initially far apart and do not perturb the chains, as shown in Fig.~\ref{fig:piston}(b) top. In this dilute, confined solution of polymers, the chain conformations are well-described by the blob model; each chain is a series of blobs, stretched along the long axis of the cylinder without large-scale backfolding~\citep{deGennesBook, Pincus76}. The origin of this \emph{linear} ordering of confined chain is entropic because overlapping blobs costs free energy~\citep{Jun07, Grosberg82}. As we shall discuss below, it can explain the principal linear chromosome organization observed in rod-shaped bacteria.

Here, the notion of ``blob'' is the central concept of our approach to bacterial chromosomes. It can formally be defined as an imaginary sphere within which the local chain segment densities are correlated~\citep{DoiEdwards}. In the above-mentioned case of dilute solution in a cylinder, the blob size $\xi$ is defined by the width of the cylinder $D$. It is important to note that a blob is also a unit of free energy of the chain and each blob contributes $\sim 1~k_BT$ to the free energy, regardless of its size~\citep{Jun07}. As we shall discuss in Sec.~\ref{sec:discussion}, for bacterial chromosomes, the closest concept to the polymer blobs in the literature is macrodomains~\citep{Espeli06} or ``topological domains''~\citep{Higgins05}. For instance, the latter denotes the largest independent chromosomal subunits, whose topological changes (such as cutting) do not affect other neighboring domains of the chromosome (for example, one may imagine loops that are independently anchored to one another). Indeed, it is also our view that chromosomes are best understood as a series of blobs, where each blob acts as an independent structural unit [see, also, \citet{Oleg07}]. 

If we compress the pistons further, the two chains will touch one another and then partly overlap. This, however, is not a favorable situation because overlapping blobs costs conformational entropy and, thus, the two chains repel each other\footnote{This chain demixing can be understood intuitively. If we freeze the red chain, its inner envelope volume can be considered as a system of fixed obstacles. Thus, the blue chain moves away from the red chain, because the blue chain's gain in conformational entropy exceeds its entropic loss by occupying smaller volume free of the obstacles. This phenomenon is due to the chain connectivity.}~\citep{Kleckner04, BatesKleckner05, Jun06} and retract to occupy each half of the cylinder. Since the total accessible volume within the cylinder has decreased by the moving pistons, the total internal energy has to increase. In other words, the blob size will decreases accordingly (see above) [Fig.~\ref{fig:piston}(b) upper box].

We emphasize that the principal organization of the retracted chains is still linear, and this average linear ordering is maintained as long as the chains stay segregated. If the piston continues to move and increase the compression of the chains, the blobs also continue to become smaller to reflect the increased free energy. Once the blobs reach their critical size, the chains gain more conformational entropy by \emph{mixing} with each other, and the chains lose their principal linear ordering and instead become a random walk of the connected blobs (Fig.~\ref{fig:piston}(b) lower box).\\

\subsection{Phase diagram for mixing and de-mixing of polymer solution in a closed geometry.} 
\label{subsec:phasediagram}
\begin{figure*}[tb]
 \centering
\includegraphics[width=18cm]{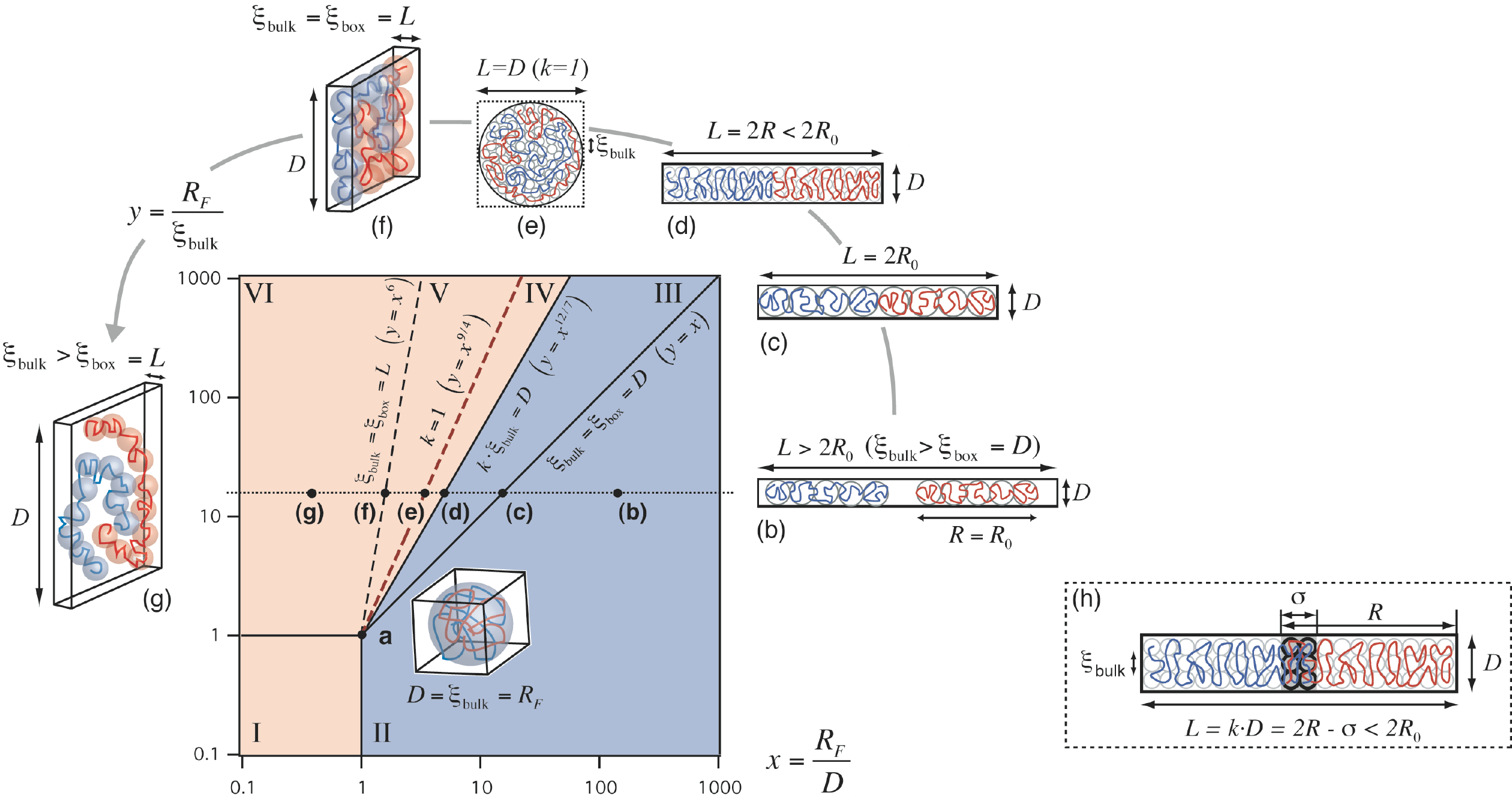}
 \caption{Phase diagram of a 2-chain polymer solution confined in a closed geometry of dimensions $D$$\times$$D$$\times$$L$. The x-axis represents the shape of the geometry. The y-axis represents the concentration of the polymer solution. Regimes with a red background represent mixed states, while those with a blue background represent segregated states. The dotted horizontal line shows the effect of aspect ratio of confinement on two chains of constant length $N$. Shown along this line are a few representative chain conformations (b)-(g). For cubic confinement with side $D=R_F$ (a), two chains mix at the free-energy cost of order $k_BT$~\citep{Grosberg82, Jun07}. (h) Same as (d), but the two chains are in partial overlap of penetration depth $\sigma$. Due to compression, the correlation length $\xi_\mathrm{bulk}$ of the confined chains is smaller than the width of the rectangular box $D$ (see the stacking blobs). [$R_\mathrm{0}$ is the equilibrium end-to-end distance of an unperturbed chain in a long cylinder, and $k$ the aspect ratio of the box.] See Appendix~\ref{appendix:phasediagram} for more technical details.}
 \label{fig:phasediagram}
\end{figure*}

The piston analogy illustrates how the physical properties of the chromosomes interplay with the confining geometry and its volume. We have obtained by rigorous analysis the phase diagram in Fig.~\ref{fig:phasediagram}, which describes the organizations of two chains in a rectangular box characterized by two length scales, width $D$ and length $L$ of the box.

The dimensionless quantity $x=R_F/D$ is the ratio between the size of the unperturbed chain ($R_F$) and the width of the cylinder; thus, the $x$-axis represents how strong the confinement is. The $y$-axis indicates how dense the polymer solution is in the box, because the correlation length (blob size), $\xi_\mathrm{bulk}$, decreases monotonically as a function of the monomer density of the polymer in solution (see Appendix~\ref{appendix:phasediagram}).

Our phase diagram is divided into two regions, one where chains segregate (blue), the other where they mix (red) (Fig.~\ref{fig:phasediagram}). The diagram is further divided into six regimes. The boundary between segregation and mixing is given by the following geometric condition
\begin{equation}
 \label{eq:boundary}
 k= \frac{L}{D} = \frac{D}{\xi_\mathrm{bulk}},
\end{equation}
\i.e., this is where the aspect ratio of the box, $k$, equals the ratio between the width of the box and the size of the blobs. The rest of the phase diagram can be best understood as we move along the horizontal dotted line in Fig.~\ref{fig:phasediagram}. Since $\xi_\mathrm{bulk}$ depends only on the chain density, for chains with fixed contour length, this line represents the change of the aspect ratio of the confining box ($k$) from filamentous ($k > 1$) to slab-like ($k < 1$) through cubic/spherical ($k = 1$) shapes, keeping constant the volume of the box. Note that the longer the box is the better the chains segregate, and that the chains can readily mix in a sphere ($k = 1$)~\citep{Jun07}.

We leave a more complete description and technical details of the phase diagram to Appendix~\ref{appendix:phasediagram}.\\

\begin{figure}[tb]
 \centering
 \includegraphics[width=8.6cm]{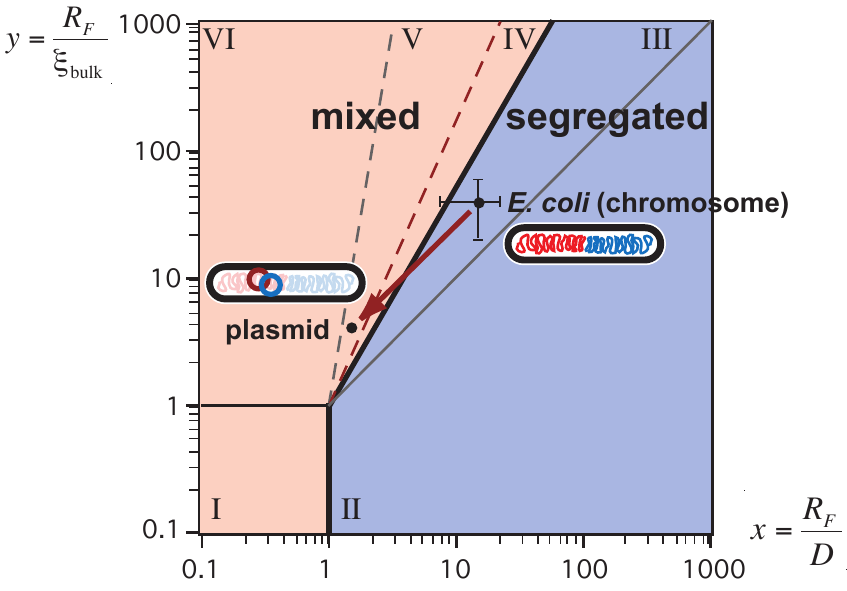}
 \caption{Segregation of DNA in \emph{E. coli}.  \emph{E. coli} is in the compressed-cigar regime III (the bars represent $50\%$ range of the average values $x = 13.6$ and $y = 37.5$, which we added to show the robustness of our estimate). Thus, the duplicated chromosomes have principal linear organization in the cell and segregate spontaneously to maximize their conformational entropy. On the other hand, plasmids are typically orders of magnitude smaller than the chromosomes, \i.e., the point for chromosome should move on the phase diagram parallel to the $y=x$ line and enters the mixing regime (e.g., V). Thus, during the course of evolution, plasmids may have acquired an additional, more active segregation mechanism other than entropic repulsion to ensure each daughter cell receives at least one copy.}
 \label{fig:plasmid}
\end{figure}

\subsection{Application to bacterial chromosomes.} 
\label{subsec:chromosome}

Using the phase diagram presented in Fig.~\ref{fig:phasediagram}, we can predict whether a specific organism can segregate its chromosomes using only conformational entropy -- with no active control mechanism. To this end, we examined the model system \emph{E. coli} B/r (H266), one of the most well-studied and well-documented bacterial strains~\citep{WoldringhOdijk}. When in steady state, slowly growing (doubling time of 150 minutes) \emph{E. coli} of this strain have a rod-shaped cell with hemispherical caps, with average length is 2.5$\mu$m and the width 0.5$\mu$m. However, the actual dimension in which chromosomes are confined inside a cell, namely the nucleoid, is smaller because of the well-known phenomena of nucleoid compaction~\citep{WoldringhOdijk}. As a result, these values are estimated to be $D = 0.24 \mu$m and $L_0 = 1.39 \mu$m for new-born cells containing a single chromosome\footnote{The population average is $\bar{L} = 1.8 \mu$m~\citep{WoldringhOdijk} with $L_0 \approx 0.7 \bar{L}$~\citep{Kubitschek81}, and we assumed a linear relationship between cell size and nucleoid length.}. 

On the other hand, the correlation length $\xi_\mathrm{bulk}$ is a more difficult quantity to measure experimentally, because a number of factors contribute to chromosome organization in bacteria (e.g., molecular crowding, supercoiling and DNA-protein interactions)~\citep{Stavans06}. Nevertheless, we can safely assume that the lower bound of $\xi_\mathrm{bulk}$ should be the persistence length of dsDNA, $\ell_p = 50$nm.\footnote{For supercoiled DNA, the persistence length of the plectoneme is estimated to be twice that of dsDNA~\cite{MarkoSiggia95}.} Note that recent experimental studies have shown that there are as many as $\simeq 400$ topological domains in \emph{E. coli} chromosome, corresponding $\simeq 10$kb per domain~\citep{Higgins05}. If we assume that each domain occupies $d^3$ of a nucleoid volume, we obtain $d \simeq 53$nm, about the same as $\ell_p$ of bare DNA. 

Very recently, Krichevsky and colleagues have measured $\xi_\mathrm{bulk}$ of nucleoid isolated from the same \emph{E. coli} strain [B/r (H266)] using a more direct method of fluorescence correlation spectroscopy (FCS)~\citep{Oleg07}. Their careful analysis of the amplitude of the FCS correlation functions of randomly fluorescent-labeled nucleoids revealed $\approx$50kb of structural unit, which corresponds to $d \simeq 87$nm of the B/r strains \emph{in vivo} [based on the nucleoid volume $\approx$0.06$\mu m^3$~\citep{WoldringhOdijk}]. They also calculated the physical size of the unit based on less direct measurement of the diffusion constant of isolated nucleoids. Their estimated value was $d = 70 \pm 20$nm, again, in good agreement with $d = 87$nm above. Indeed, we have used these parameters including the average size of the isolated nucleoids $R_F = 3.3 \mu$m~\citep{Oleg07} (from the average spherical volume of natively supercoiled nucleoid, 18 $\mu$m$^3$) and obtained $x = 13.6$ and $y = 37.5$ -- the \emph{E. coli} B/r strains are in the segregation regime II (Fig.~\ref{fig:plasmid}). Thus, the replicating chromosomes could segregate purely driven by conformational entropy, and they will remain demixed~\citep{BatesKleckner05, Jun06}.\\

\subsection{Application to plasmid segregation -- evolutionary necessity for segregation strategies.} 
\label{subsec:plasmid}

Another important biological question concerns plasmid segregation in bacteria. Importantly, most low-copy number plasmids, such as R1, are believed to rely on more biological, protein-based active segregation mechanisms [e.g., \emph{par} system~\citep{mullins07, schumacher07}]. Our phase diagram can explain the need for dedicated segregation mechanisms as follows: Since the size of typical plasmids is $\sim$10kb, they can be considered as small ``chromosomes'' from a physical point of view (note: typical chromosomes are $\sim$$10^3$kb).  At the insertion of two such plasmids in bacteria, the total amount of DNA (and, thus, $\xi_\mathrm{bulk}$) remains practically constant, although the total number of chains increases to $n=4$. In our phase diagram, this corresponds to changing both $x$ and $y$ of chromosome by a factor of $\sim 10^1$ (since $R_F \sim N^{3/5}$), {\i.e.}, the plasmids enter the mixing regime V~\footnote{
Note that, in principle, there are four determining parameters of our phase diagram, where the fourth parameter is the number of chains $n$. If we release the constraint imposed on $n$ [e.g., $n=2$ (chromosomes) and $n=4$ (chromosomes + plasmids)], while keeping constant $D$ and $\xi_\mathrm{bulk}$, the two cells are connected by a diagonal line in the phase diagram because, then, $R_F$ is the only additional variable (Fig.~\ref{fig:plasmid}).} 
and they will distribute randomly inside the cell. Indeed, recent experimental results on the mobility and distribution of synthetic minimalized RK2 plasmid lacking the partitioning system are fully consistent with our prediction of random distributions~\citep{Pogliano08}.

The quantitative reasoning above has broader implications for segregation strategies and copy-number control of plamids~\citep{Nordstrom03}: (i) An \emph{active} segregation mechanism, especially, for low-copy number plasmids~\citep{Gerdes04, Adachi06, Pogliano08}, or (ii) ``Random'' segregation, \i.e., without a protein-based segregation mechanism, plasmids may produce multiple copies so that, despite their small sizes, their chance of survival increases upon cell division. In random segregation, copy-number control is an interesting and important issue [See, for example, \citet{Tomizawa} and references therein].\\

\subsection{Segregation of replicating nucleoid.} 
\label{subsec:shell}
\begin{figure*}[tb]
 \centering
\includegraphics[width=17cm]{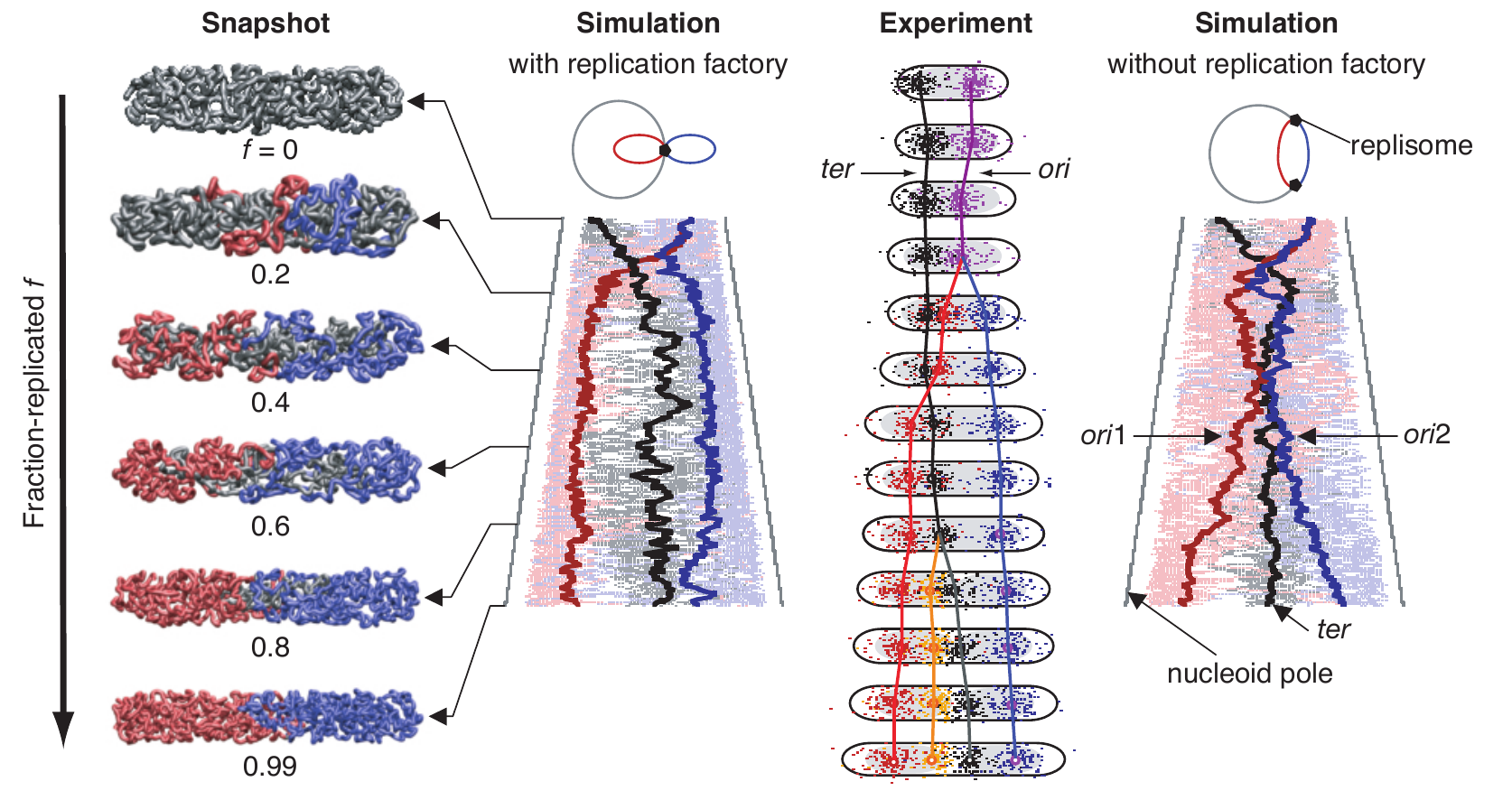}
 \caption{Chromosome segregation in \emph{E. coli}: simulation vs. experiment. (Left) A series of typical conformations of a replicating circular chain from 0\% to 99\% replication (un-replicated strand in gray, replicated strands in red and blue). We also present two sets of segregation pathways (\emph{ori-ter} trajectories during replication) with and without ``replication factory''; the third set of simulation where \emph{ori} starts at the midcell can be found in Supporting Information in \citet{Jun07}. We simulated replication factories by enforcing physical proximity of the two replisomes during replication, but we did not fix their position within the cell. The dotted lines show the results of 10 individual simulations, and the thick solid lines show the average trajectories of \emph{ori} (red and blue) and \emph{ter} (black). (Center-to-Right) We juxtapose the simulations with the published data in~\citet{BatesKleckner05} in an attempt to capture the main features of the experimental observations. For comparison, we used the fraction-replicated $f$ as our ``universal clock'' and scaled the height of the simulation trajectories to match $0 < f < 1$ range of the data. [Reprinted from Fig. 5 of \citet{Jun06}. Copyright (2006) National Academy of Sciences, U.S.A.]}
 \label{fig:ecoli}
\end{figure*}

So far, we have discussed the physical conditions in which duplicated chromosomes will stay segregate, \i.e., the conditions for a primordial driving force for chromosome segregation. How then can we explain observed trajectories?

\begin{figure}[tb]
 \centering
 \includegraphics[width=8.6cm]{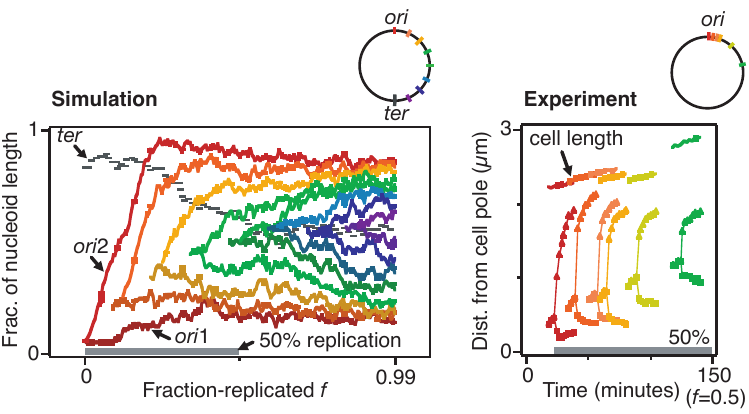}
 \caption{Chromosome segregation of \emph{C. crescentus}, comparing simulation (Left) vs. experiment~\citet{Viollier04} (Right). The simulated trajectories are the average of 26 individual simulation runs (or ÔÔcellsÕÕ); we show the trajectories of nine representative loci (including \emph{ori} and \emph{ter}) on the right-arc of a circular chromosome for the entire duration of replication (up to 99.9\%), whereas experimental data are only available for trajectories up to 50\% of replication. For clarity, we show the trajectories only from the onset of replication of each locus. A full trajectory of \emph{ter} is shown, however, to emphasize its slow drift from the cell pole to the cell center during replication, in contrast to the fast, directed diffusion of \emph{ori2} in the nucleoid periphery. A main difference from the \emph{E.coli} simulation is the additional assumption that we kept \emph{ori1} in the volume near the stalked pole until 10\%Ð-20\% of the chain has been replicated. The concentric-shell model used in the simulation is formal and is not inconsistent with other additional mechanisms that may act on the directed movement of \emph{ori2}, although in our simulation we did not need any such assumptions.
[Reprinted from Fig. 6 of \citet{Jun06}. Copyright (2006) National Academy of Sciences, U.S.A.]}
 \label{fig:caulobacter}
\end{figure}

Figures.~\ref{fig:ecoli} and~\ref{fig:caulobacter} show experimental data of the average positions of the chromosome loci in \emph{E. coli}~\cite{BatesKleckner05} and \emph{C. crescentus}~\cite{Viollier04}. In \emph{E. coli}, for instance, \emph{ori} first moves towards the midcell position and, then, replication starts. The duplicated \emph{ori}'s split and move,\footnote{There is an ongoing debate about how long the duplicated \emph{ori} and other loci stay together before splitting during chromosome segregation (``cohesion''). See, for example,~\citet{Hiraga01, Nanninga02, BatesKleckner05, Nielsen06}.} on \emph{average}, towards the 1/4 and 3/4 positions. In the mean time, \emph{ter} drifts slowly from the cell pole towards the midcell, crossing one of the \emph{ori} trajectories.

In \emph{C. crescentus}, loci trajectories seem even more striking~\cite{Viollier04}: one \emph{ori} stays at the cell pole and the duplicated \emph{ori} moves much faster than the growth rate of the cell (Fig.~\ref{fig:caulobacter}). Indeed, this large difference between the cell growth rate and the speed of \emph{ori} is the evidence against Jacob and colleagues' model mentioned earlier, at least, in \emph{C. crescentus}. The rest of the loci follow similar trajectories, creating a mirror-like image of the organization of duplicated circular chromosomes. \emph{Ori} is found at the old cell pole and \emph{ter} at the new cell pole; importantly, there is a one-to-one correspondence between the clock position of the chromosomal loci and their average physical locations along the long axis of the cell. In \emph{E.coli}, recent data from several labs [e.g., \cite{Wang06, Nielsen06}] also suggest similar principal linear organization, except that their clock positions are rotated by 90 degrees with the two chromosome arms between \emph{ori} and \emph{ter} occupying each cell half.

\begin{figure}[tb]
 \centering
 \includegraphics[width=7.5cm]{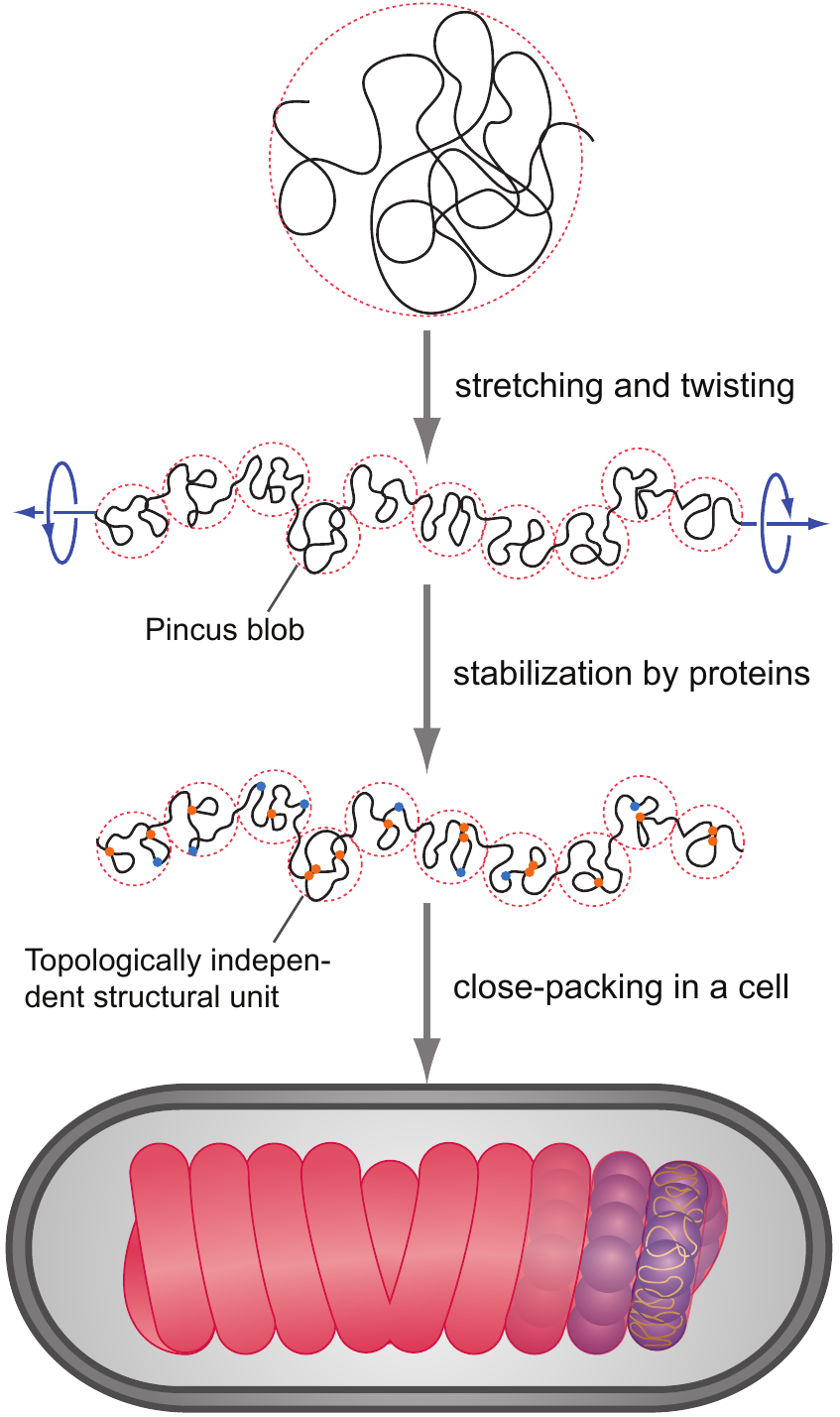}
 \caption{Illustration of chromosome as a close packing of a string of Pincus blobs (see, also, Appendix.~\ref{appendix:pincus}). The blobs schematically represent correlation-length structural units of the chromosome, which can be formed by stretching (and twisting) of the chain. (Although not essential to our theory presented in this article, independence of the topological states of individual blobs may be achieved by nucleoid-associated proteins.) To maintain an average linear organization, we envision the blobs forming a helical coil that fills the nucleoid volume. If the helicity reverses, nucleoid envelope may develop constricted, bi-lobed or multi-lobed shape. Molecular crowding and possible interactions between the inner cell-wall membrane and dsDNA protruding from the nucleoid is not shown for clarity. The gap between the nucleoid and the inner cell-wall membrane is not drawn to scale and may vary from organism to organism.}
 \label{fig:renate}
\end{figure}

To explain these experimental data, we have previously proposed the concentric-shell model~\citep{Jun06}, which was inspired by the observation of nucleoid compaction in \emph{E. coli} (see, also, Fig.~\ref{fig:renate}). Here, our prediction was that the newly synthesized DNA will move much faster in the periphery of the nucleoid near the cell-wall membrane than inside the nucleoid body (which is a meshwork of chromosomal DNA), faster than the typical timescale of the cell cycle of \emph{E. coli} ($>$ 20 min.) or \emph{C. crescentus} [$\approx$ 240 min. in~\citet{Viollier04}]. Indeed, this minimal model and assumption reproduced most of the major features of the experimental data in both organisms.\footnote{In our previous study~\citep{Jun06}, this gap was even smaller than the width of the chain in our simulations.} For \emph{E. coli}, we note that \citet{Fan07} have also used a free energy-driven string model to explain the experimental data. The most important result in their study is that the size of the chromosomal domain is important (the larger the domain is, the better chromosomes segregation), which we have explained above using our phase diagram.

We emphasize that the concentric-shell model is consistent with other models, as long as they also imply the preferential occupation of the nucleoid periphery volume by newly synthesized DNA in the early stage of cell cycle. These models include the transertion model by~\citet{Woldringh02}, which assumes an interaction between the nucleoid and the inner-cell wall membrane via co-transcriptional translation and protein translocation, or even the hypothetical role of \emph{Par} proteins on \emph{ori} transportation in the cytoplasmic space, namely, outside the nucleoid volume~\citep{Viollier04}. We believe timelapse experiments with higher spatiotemporal resolution will reveal the nature of directed motion between diffusive, biased random walk vs. transportation by polymerization or motor proteins.

\section{Discussion and frequently asked questions}
\label{sec:discussion}

Based on the piston model of the bacterial cell, we have presented the above physical biology model of bacterial chromosomes. Importantly, our model makes an experimentally testable predictions whether duplicated chromosomes will mix or segregate, and we have critically examined our predictions against the published data of \emph{E. coli} B/r strain.

We emphasize the importance of the concept of correlation length $\xi$, a central parameter of our chromosome model. To recapitulate, this is the size of the chromosomal structural unit within which the density fluctuations (and, loosely, the motions) of the chain segments are correlated. The major assumption underlying our model is that, although a real bacterial chromosome is a very complex entity because of supercoiling and the presence of various nucleoid-associated proteins (whose roles are still largely unknown), there is a single length scale $\xi$ that characterizes the nucleoid. Then, the chromosome can be interpreted as an entity consisting of these structural units, and chromosome organization in bacteria can be understood as a close packing of a string of Pincus blobs. We illustrate this view in Fig.~\ref{fig:renate} and leave the scaling analysis for equivalence between the Pincus chain \citep{Pincus76} and the compressed chain in Appendix.~\ref{appendix:pincus}.

Based on our results discussed in the previous sections, there are two general principles in understanding chromosome organization and segregation in bacteria.

	%    define "Lcount" as a counter
\newcounter{Lcount}
	%    set the "default" label to print counter as a Roman numeral 
\begin{list}{\Roman{Lcount}.}
	%    inform the list command to use this counter
{\usecounter{Lcount}
	%    set rightmargin equal to leftmargin
\setlength{\rightmargin}{\leftmargin}}
\item{The larger the correlation length is, the better chromosomes segregate.}
\item{The larger the aspect ratio is (\i.e., the longer and/or the narrower the cell is), the more chromosomes tend to demix.}
\end{list}

\noindent Below, we discuss various questions concerning our chromosome model.\\

\subsection{Role of proteins}
\label{subsec:proteins}

Proteins directly and indirectly change the physical properties of chromosomes and the cell morphology and, thus, they are important for chromosome organization and segregation in the following three contexts.\\

%\newcounter{Lcount2}
%\begin{list}{(\roman{Lcount2})}
%{\usecounter{Lcount2}
%\setlength{\rightmargin}{\leftmargin}}
%\setlength{\leftmargin}{20pt}
%\setlength{\rightmargin}{0pt}}

\noindent\emph{Correlation length of chromosome}. Various SMC or nucleoid-associated proteins such as \emph{MukBEF, HU, H-NS, IHF} and gyrase may change the size of the chromosomal structural unit and, thus, can help ensure successful chromosome segregation~\citep{Stavans06}. From our point of view, the best evidence has been given by~\citet{austin00} [see, also, \citet{Cozzarelli00} and \citet{Dasgupta01}]. These authors have demonstrated that severity of disruption of chromosome segregation of $muk^-$ \emph{E.coli} can be controlled by the level of supercoiling in the cell. Thus, neither class of proteins is the dedicated segregation machinery, but their role can be understood as, within the theoretical framework presented here, to increase the effective correlation length of the chromosome above the critical size (Eq.~\ref{eq:boundary}), which is sufficient for proper segregation  (see I above).

We also note that, because of supercoiling, a more realistic coarse-grained topology of a bacterial chromosome is that of the branched polymer rather than a purely linear chain. This branched structure will enhance the tendency of chromosome pushing even more~\citep{marko98, Jun06, Vilgis00}.\\

\noindent\emph{Cell shape,  \emph{MreB}, and bacterial ``cytoskeletal'' proteins}. \emph{MreB} and other proteins~\citep{Gerdes04} are important to maintain the high aspect ratio of the cell (see II above). It is our contention that, if these proteins contribute to chromosome segregation, it is about changes in cell shape as our model suggests.\\

\noindent\emph{Cell growth and division.} Since the cell size (relative to the correlation length) is also important, proteins that regulate cell growth and division are also important to ensure that the cell reaches the appropriate size or mass for proper segregation~\citep{AnneLevin07}. In this context, the length-measuring ``devices" such as \emph{MinCDE} are also important, and proteins involved in cell division such as \emph{FtsZ} are also relevant [see~\citep{Lutkenhaus07} and references therein], if cell constriction should help resolve partially overlapping nucleoids~\citep{Woldringh99}.

%\end{list}

\subsection{Shouldn't a hypothetical motor protein be enough to segregate chromosomes in bacteria?}
\label{subsec:motor}

No. There is increasing experimental evidence that \emph{ori} and perhaps some other chromosomal loci are localized or tethered at specific intracellular positions [see, for example, \citet{Stavans06} and references therein]. However, one still must explain how the hypothetical transportation of a small fraction (e.g., \emph{ori}) of millions of basepairs of DNA from one position inside the cell to another can dictate, if any at all, the directed movement, segregation and organization of the rest of the chromosome. These observations thus require understanding of more basic physical principles (e.g., Fig.~\ref{fig:phasediagram}).

We note that there are special cases where motor proteins are indeed involved in transportation of DNA from one position to another inside a bacterial cell. The examples include \emph{SpoIIIE} for sporulation in $B. subtilis$ and \emph{FtsK} for dimer resolution and other ``rescue'' tasks in \emph{E.coli} [\citet{Barre07} and references therein]. However, it is important to realize that these proteins \emph{translocate} DNA, \i.e., they take advantage of the directionality provided by the septum, which is entirely different from segregation of replicating chromosomes.\\

\subsection{What is the timescale?}
\label{subsec:timescale}

In the segregation regime, segregation is a \emph{driven} process and two intermingled chains drift at typical time scale of $\tau \sim N^2$~\citep{Jun06, arnold07a}, which is much faster than diffusion (reptation) timescale of $\tau \sim N^3$ in the channel. Timescale of polymer motions in the presence of confinement involves severe finite-size effects~\citep{arnold07b}, and it is far more difficult to predict the timescale for segregation of \emph{duplicating} chains. For the concentric-shell model in our simulations for \emph{E.coli} and \emph{C. crescentus}, we assumed a separation of timescale that the relaxation time of the replicating segment of DNA is much smaller than the typical timescale of cell cycle (Sec.~\ref{subsec:shell}). Such fast movements of DNA segment have indeed been reported in recent experiments, that the 15kb minimalized plasmid RK2 explores about $50\%$ of the cell volume only within 2 minutes~\citep{Pogliano08}.

We do feel that the issue of timescale can be addressed adequately only via experiments using novel physical techniques.\\

\subsection{How about topologically catenated chromosomes?}
\label{subsec:catenation}

Imagine two topologically concatenated ring polymers confined in a rectangular box, where our phase diagram in Fig.~\ref{fig:phasediagram} predicts segregation of two linear chains of the same contour length as the rings. These two ring polymers will still segregate to occupy each half of the box, \i.e., the topology of the system localizes on average at the center of the box (unpublished results). It is an intriguing question whether this localization of chain topology, which is a \emph{global} property of the system, will influence the action of topoisomerases, which have access to only \emph{local} information. We speculate that the entropy-driven directionality of the movements and relative orientation of the confined chains will, at least in part, help topoisomerases decatenate the replicated chains before cell division.\\

\begin{figure}[tb]
 \centering
 \includegraphics[width=8.6cm]{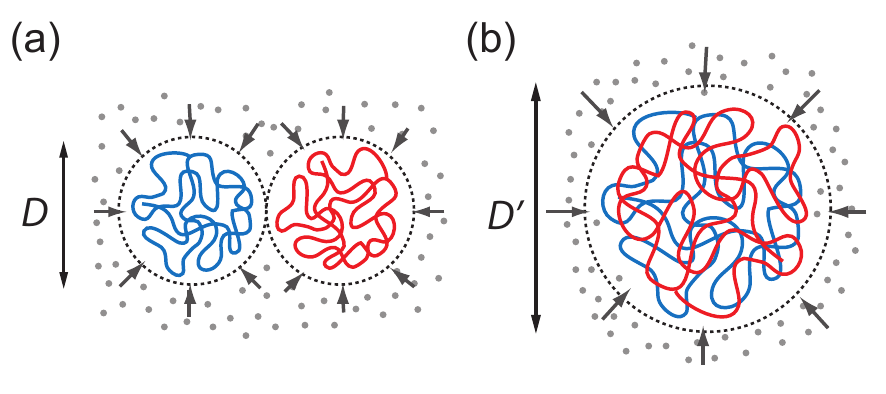}
 \caption{Two possible scenarios of the role of molecular crowding on chain interactions. (a) segregation (b) mixing. Each chain consists of $N$ monomers. Calculations show that crowding, in the first approximation, should not influence mixing vs. de-mixing of the chains (see text).}
 \label{fig:crowding}
\end{figure}

\subsection{Bacterial cells are very crowded: Will molecular crowding influence chromosome organization and segregation?}
\label{subsec:crowding}

In the first approximation, no. One may foresee two opposing arguments: (A) As the chains are compressed by depletion effect, the inner space occupied by the chains has higher density of polymers, \i.e., less accessible volume by the other chain and, thus, the tendency of de-mixing will increase because of squeezing, or (B) Molecular crowding exerts an effective osmotic pressure, which will make the chains mix.

In fact, a simple calculation suggests that there is a balance between these two effects. To see this, let us consider the free energy of the compressed chains in Fig.~\ref{fig:crowding} as follows~\citep{GrosbergBook, Sakaue06, Cacciuto06, Jun07}.
\begin{eqnarray}
 \label{eq:crowding}
 \beta \mathcal{F}_\mathrm{A} &=& 2C\bigg( \frac{N^\nu}{D} \bigg)^\frac{3}{3\nu-1}\\
 \beta \mathcal{F}_\mathrm{B} &=& C\bigg[ \frac{(2N)^\nu}{D^\prime} \bigg]^\frac{3}{3\nu-1},
\end{eqnarray}

\noindent where $C$ is the prefactor. These two free energies for the compressed chains become equal at $D^\prime = 2^{1/3} D$, \i.e., when their total volumes are the same, $V_A = V_B$. Thus, compression due to depletion by itself will not influence mixing vs. de-mixing of the chains, even if it could bring the chains together and change their envelope shape (surface tension).\footnote{The origin of this result is due to the form of the free energy in Eq.~\ref{eq:crowding}~\citep{Grosberg82, Sakaue06, Jun07}, which has been constructed to be a function of only the monomer density. In other words, regardless of the shape of the envelope surrounding the chains, the free energies will be the same as long as the volumes are also the same, and vice versa. Additional consideration of the interactions at the surface will change the envelope shape, but not segregation/mixing.} Computer simulation results also support the neutral effect of molecular crowding (Axel Arnold, personal communication). 

On the other hand, crowding can influence the local organization of chromosome. For instance, looping is one way to achieve compaction of chromosomes, and the entropy gain by depletion attraction between DNA segments can be larger than the entropy loss by DNA looping \citep{Cook2006}.

\subsection{Bacteria exist in various cell shapes and composition of chromosomes. Are there general strategies for successful chromosome segregation in bacteria?}
\label{subsec:strategy}

\begin{figure}[tb]
 \centering
 \includegraphics[width=8.6cm]{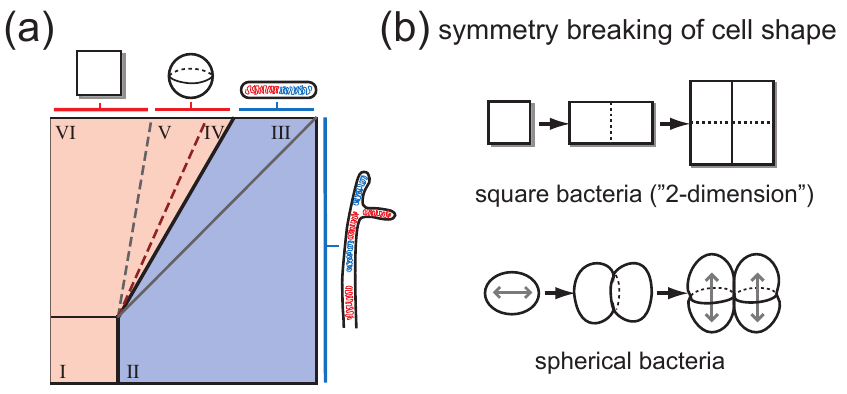}
 \caption{Possible cell-shape-dependent strategies for chromosome segregation. (a) According to our phase diagram, filamentous and rod-shaped cells are most favored by entropy-driven segregation. (b) Thin, square bacteria such as Walsby's square bacterium change their direction of growth after every division (typically they are found in composition of 1x1, 2x2, 4x4, 8x8 cells and so on).  On the other hand, round cells divide at alternating perpendicular planes. The arrows in the round cells indicate oscillation of the MinCDE proteins along the long axis of the ellipsoidal cells, which help determine the mid plane of the cell.}
 \label{fig:strategy}
\end{figure}

Since chromosome segregation is one of the defining processes of any cell, its basic mechanism must have been conserved across branches of life and organisms of diverse shapes. Based on our phase diagram (Fig.~\ref{fig:phasediagram}) and the two general rules $I~\&~II$ presented above, we can speculate about how bacteria may create favorable physical conditions for partitioning the chromosomes (Fig.~\ref{fig:strategy}).\\

\noindent {\it Filamentous bacteria.} We propose an entropy-driven,``random'' segregation mechanism for filamentous bacteria. As we have shown both analytically and numerically, polymers confined in a narrow cylindrical geometry strongly resist overlap and, thus, repel one another~\citep{Jun06, arnold07a}, where the timescale of disentanglement of overlapping polymers is much shorter than that of pure diffusion. Indeed, recent experimental study on the filamentous cyanobacterium \emph{Anabaena sp. PCC 7120} has revealed Gaussian distribution of DNA content in each daughter cells after the septum formation~\citep{hu07}. Importantly, the variation of the distribution was much larger than a value expected if the two daughter cells were identical, suggesting a random event involved in chromosome segregation. Also, \emph{Anabaena sp. PCC 7120}, like many other filamentous bacteria, is polyploid and contains multicopy of chromosomes per cell, $\sim$ 10 \citep{hu07}, which seems to moderate the effect of random segregation.  Since chromosomes occupy much larger volume than plasmids inside the cell, this apparent random segregation process is sufficient for the viability of the cell as long as multiple copies of chromosomes are produced before division.\\

\noindent {\it Rod-shaped bacteria.} We have critically examined the model system \emph{E. coli} B/r strain and explained why duplicated chromosomes will not mix. Thus, other organisms of similar cell shape and volume must benefit from the entropy-driven driving force of segregation, with or without any organism-specific segregation mechanisms, as long as the the correlation of chromosome is also comparable to that of \emph{E. coli} ($\approx 100$ nm).\\

\noindent {\it Square bacteria.} Perhaps one of the most striking bacterial shapes is that of Walsby's square bacterium~\citep{Walsby80}. This organism has a very thin and square-shaped cell ($\approx100$nm thick and $\approx 2 \mu$m wide with aspect ratio 1), a ``2-dimensional'' creature resembling a postage stamp. As we have explained above, polymers in highly symmetric confinement can readily intermingle. In other words, entropy-favored condition for segregation can be created when the cell breaks its symmetry during growth or division. Indeed, microscopy images so far suggest that that square bacteria grow and divide at alternating perpendicular axes and planes, respectively [Fig.~\ref{fig:strategy}(b)]. Also, in a thin slab (\i.e., 2D confinement), using the approach taken in~\citet{Jun07}, it is straightforward to show that polymers repel much more strongly than in bulk 3D.\footnote{The free-energy cost of chain overlapping crosses over from $\beta\mathcal{F} \sim n^{d\nu_d/(d\nu_d - 1)}$ to $\sim n^{d/(d-2)}$, where $\nu_d = 3/(d+2)$ is the Flory exponent and $d$ the spatial dimension. [\citet{Jun07}; B.-Y. Ha, personal communication]} (Note that the cells are typically observed in 2x2, 4x4, 8x8 stamp-like configurations.) Therefore, we believe that simple symmetry breaking by cell growth and the thin 2D geometry suffice to separate the chromosomes entropically.\\

\noindent {\it Spherical bacteria.} The perfect symmetry of the cell shape means that the confined chains do not have any preferred conformations between mixing and de-mixing, although their global reorganization could readily be achieved~\citep{Jun07}. Since little data is available on chromosome organization in spherical bacteria, we can only speculate about possible contributing factors to segregation. At the chromosome level, supercoiling-induced branched structures of the chromosome will increase the tendency of demixing~\citep{marko98, Jun06, Vilgis00}, perhaps helped by nucleoid-associated proteins to increase the correlation length (rule I above). More importantly, at the cell level, symmetry breaking of the cell shape and invagination of the cell during division could help resolve partially intermingled chromosomes [Fig.~\ref{fig:strategy}(b)]. Indeed, real cells are never perfectly spherical, and \citet{Huang04} have shown numerically that Min-protein oscillations can be achieved along the long axis of the \emph{nearly} round cell, where the equatorial radii differ by as small as 5\%. This may explain the observed division of round cells at alternating perpendicular planes~\citep{Margolin02}. Along with the case of the filamentous bacteria discussed above, it is tempting to speculate that chromosome segregation in spherical bacteria is also a ``random'' process and, thus, polyploidy can be a strategy to increase the chance of successful distribution of chromosomes to daughter cells. For example, the endosymbiotic bacteria of aphids, \emph{Buchnera}, which cannot divide by itself outside the host eukaryotic cell, contains over 100 copies of genome~\citep{Ishikawa99}.\\

\begin{figure}[tb]
 \centering
 \includegraphics[width=8.0cm]{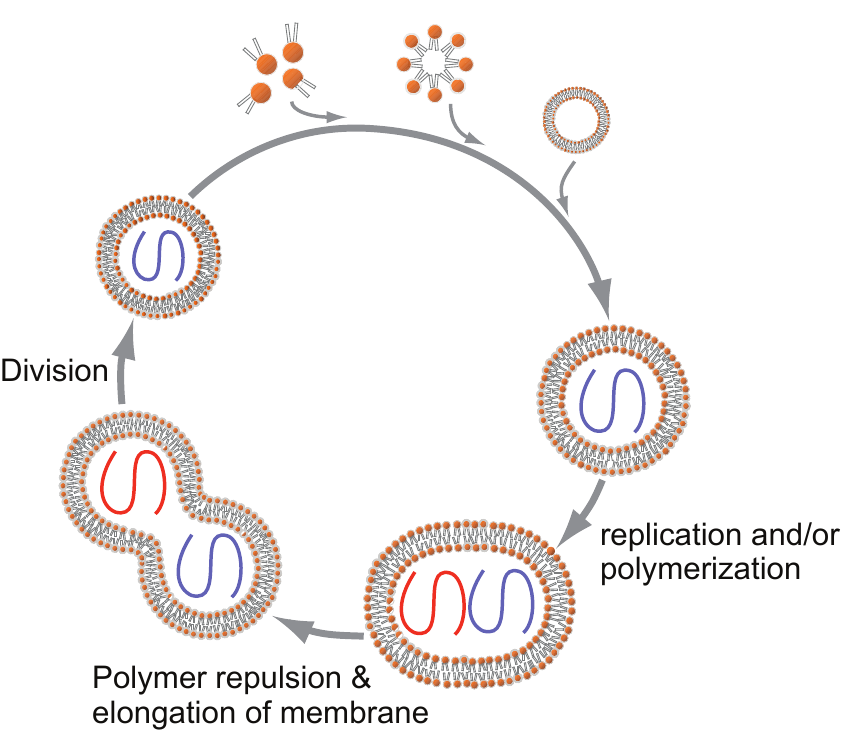}
 \caption{Entropy-driven segregation of encapsulated polymers and its potential role on the cell cycle of a protocell. Repulsion between the replicated chains can break the spherical symmetry of the vesicle, and, thus, may facilitate its spontaneous division. Moreover, the daughter vesicles containing the duplicated polymers may have similar sizes. (The three minisketches at the top represent input from environment for feeding the growth.) These scenarios could be tested both theoretically and experimentally. [Adapted from Figure 3 in ~\citet{Szostak}.]}
 \label{fig:protocell}
\end{figure}

\subsection{Early life and artificial cells}
\label{subsec:artificialcell}

Previously, we proposed that the entropy-driven segregation of polymers could be implemented in designing an artificial cell, and that it has implications on early life~\citep{Jun06}. How could an artificial cell achieve its basic cellular processes? While Szostak and colleagues have proposed a system of a self-replicating vesicle and replicases as a protocell~\citep{Szostak}, cell division mechanism as well as the size regulation of daughter cells of a protocell are far from being understood.

We suggest that two long polymers, formed by replication/ligation of smaller molecules within a spherical vesicle repel each other because of entropy. If the polymer-polymer repulsion is strong enough to break the spherical symmetry of the vesicle, this process may lead to membrane fission, preferentially at the mid-cell position, which is defined by the surface of contact between the two repelling polymers. This would further help regulate the size distribution of newly formed daughter cell as illustrated in Fig.~\ref{fig:protocell}.  Although it is in general a formidable theoretical problem to predict the shape of the closed membrane in a given environment, it would be highly desirable to obtain a phase diagram similar to Fig.~\ref{fig:phasediagram} for soft walls, and possibly test our proposed idea here experimentally using macromolecules encapsulated in vesicles or micro drops.

\section{Final remarks}
\label{sec:final}

In recent years, biology has become increasingly interdisciplinary. In particular, in the new discipline of systems biology, researchers trained in physical and other quantitative sciences are making significant contributions. From a physicist's point of view, however, it is interesting and important to notice that many of the contributions made by physicists so far have involved bringing new tools and ways of thinking to (systems) biology, rather than new understanding about the role that basic physical principles play in governing the fundamental biological processes inside the cell. But aren't biological entities also physical objects?  The membranes defining the envelope of the organelles; the polymers constituting proteins, spindles, and chromosomes -- these are physical objects and, although biophysical approach has uncovered many of their important physical properties, their biological implications need to be more fully explored.

Thus, in seeking a more direct relationship between physics and biology, with the example of bacterial chromosome segregation, I hope to have convinced the reader that there is a seamless ``dialogue" between physical and biological processes involved in the bacterial cell cycle; and that deeper understanding of the relationship between the physical principles and their biological implications may even shed new light on the major transitions in evolution~\citep{MaynardSmith}.\footnote{For instance, in the presence of multiple chromosomes in eukaryotes, the cell faces an additional challenge in chromosome segregation because each daughter cell must receive not only the correct number but also the correct \emph{set} of chromosomes. From out point of view, this strongly argues why mitosis requires more sophisticated segregation machinery as well as the checkpoints; whereas entropy might be sufficient for chromosome segregation in most bacteria.\\ 
\indent On the other hand, the hallmark of mitosis is the compact sister chromosomes of well-defined shapes being held together, waiting for the mitotic spindle to separate them. However, these chromosomes must go through a period of intermingling during replication, before mitosis. Thus, there is a process analogous to bacterial chromosome segregation, namely, \emph{de-mixing} of replicated eukaryotic sister chromatids, which we distinguish from \emph{segregation} by spindle. Although the exact mechanism of this de-mixing process has remained elusive, we believe that the basic physics behind is the same as what we presented in this article.}

\begin{acknowledgments}
This work would not have been possible without my long-term collaborations with A. Arnold, B.-Y. Ha, N. Kleckner, and C. Woldringh. I thank J. Bechhoefer, M. Brenner, R. D'Ari, D. Frenkel, M. Kardar, O. Krichevsky, P. A. Levin, R. Losick, J. Marko, B. Mulder, A. Murray, D. Nelson, B. Stern, F. Taddei and numerous other colleagues for helpful suggestions. I am also grateful to N. Kleckner for introducing the notion of ``piston'' to me, and to R. Hellmiss for the illustration of chromosome model in Fig.~\ref{fig:renate}. This article is dedicated to the memory of J. Raoul-Duval.
\end{acknowledgments}

\appendix

\section{More technical description of the phase diagram in Fig.~\ref{fig:phasediagram}.}
\label{appendix:phasediagram}

\noindent Below, we provide more technical descriptions of individual regimes with references so that the reader can reproduce the boundary conditions of the phase diagram in Fig.~\ref{fig:phasediagram}. [See, also,~\citet{Daoud77, Brochard79, Teraoka04}].\\

\noindent {\it I}. This is the ``bulk'' dilute regime, where the interchain distances are much larger than the size of the chain in confinement. When $D=L=R_F$, the correlation length $\xi_\mathrm{bulk}$ also equals the size of the chain $R_F$, and the two chains can overlap at the free-energy cost of order $k_BT$ [Fig.~\ref{fig:phasediagram}(a)]~\citep{Grosberg82, Jun07}.\\

\noindent {\it II}. Imagine dilute solution in a long cylindrical box, where there is a free space between the two chains [Fig.~\ref{fig:phasediagram}(b)]. The size of individual chains scales as $R_\mathrm{0} \sim (N/g)D \sim N D^{-2/3}$ (where $g$ is the number of monomers per blob). In case of overlapping, the chains strongly repel one another and, thus, their segregation is much faster than reptation (typical timescale of $\tau \sim N^2$ vs. $\tau \sim N^3$, respectively)~\citep{arnold07a, Jun06}. Note that $\xi_\mathrm{bulk} > D$ in this regime, whereas the blob size of the chain due to confinement is $D$. Thus, the monomer-monomer correlation length inside the box is $\xi_\mathrm{box} = D < \xi_\mathrm{bulk}$. 

If we gradually reduce the aspect ratio of the box, keeping constant the volume of the box (thus, fixing $\xi_\mathrm{bulk}$), the number of blobs of each chain decreases and eventually the two chains make contact at $L = 2R_\mathrm{0}$, {\i.e.}, $\xi_\mathrm{bulk} = \xi_\mathrm{box} = D$ ($y = x$). This situation is depicted in Fig.~\ref{fig:phasediagram}(c), at which the chains enter the semi-dilute regime inside the elongated box.\\

\noindent {\it III}. If we continue the above process of reducing the aspect ratio of the box, the two chains are gradually compressed but do not mix as long as their principal conformations are linear [Fig.~\ref{fig:phasediagram}(d).] As we discussed in Sec.~\ref{subsec:chromosome}, this is the most important regime for rod-shaped bacteria. To see the segregation of overlapping chains, let us employ the following Flory-type free energy of individual chains:
\begin{equation}
 \label{eq:rejected}
 \beta\mathcal{F}_1(R) = \frac{1}{2} \bigg[\frac{R^2}{(N/g)D^2} + \frac{D(N/g)^2}{R}\bigg],
\end{equation}
where $\beta = 1/k_BT$~\citep{JunHa07}. On the other hand, it costs $\sim k_BT$ of free energy for two blobs to significantly overlap regardless of their size~\citep{Grosberg82, Jun06, Jun07}. Thus, the total free energy of the system of two partially overlapping chains described in Fig.~\ref{fig:phasediagram}(h) can be written as
\begin{eqnarray}
 \label{eq:overlap}
 \beta \mathcal{F}_\mathrm{tot} &=& \Big(2+\frac{\sigma}{R}\Big)\beta \mathcal{F}_\mathrm{1}(R) \notag \\
 &=& \frac{k}{2} \bigg(\frac{4}{L^\prime}-\frac{1}{R^\prime}\bigg) \bigg[ {R^\prime}^2 + \frac{1}{R^\prime} \bigg],
\end{eqnarray}
where $L^\prime = L/R_0 \leq 2$, $R^\prime = R/R_0 \leq 1$, $D^\prime = D/R_0$. Since $\partial \beta \mathcal{F}_\mathrm{tot}/ \partial{R^\prime} = 3k/2 > 0$ at $R^\prime = L^\prime/2$ (namely, when each chain occupies each cell half), Eq.~\ref{eq:overlap} indeed predicts a free-energy barrier against mixing of two segregated chains. Moreover, one can show numerically that the free energy $\mathcal{F}_\mathrm{tot}$ is a monotonically increasing function of $R^\prime \geq L^\prime/2$ for $L^\prime \gtrapprox 0.8$.\footnote{Since the $N^2/R$ dependence in Eq.~\ref{eq:rejected} is an well-known overestimate of Flory-type approach for $R \ll R_0 $, the $L^\prime \approx 0.8$ is an upperbound. In other words, two chains should segregate in an even shorter box. For the same reason, we cannot use Eq.~\ref{eq:overlap} to determine the condition for crossover to regime IV. To our knowledge, a correct form of the free energy that is valid even for largely perturbed chains (namely, for all range of $R$) is not available.}\\

\noindent {\it IV, V, VI} (mixing regimes). As the aspect ratio approaches $1$, the two chains lose their linear conformations and mix with each other. The reason for this transition is that the bulk correlation length $\xi_\mathrm{bulk}$ becomes much smaller than both $D$ and $L$, and the chain conformation becomes a random walk of a string of blobs (blob size $\xi_\mathrm{bulk}$). We described the physics of this regime in detail in~\citet{Jun07}. 

At the boundary between $III$ and $IV$, transition occurs because individual chains are contracted and lose their linear ordering. The size of individual chain is that of the random-walk conformation, $R_\parallel \sim N^{1/2}\Phi^{-1/8}$ with the volume fraction of the monomer $\Phi \sim N/R_\parallel D$. Based on this, it is easy to show the boundary condition $y=x^{12/7}$~\citep{Lal97}. Since the free energy in this regime is given by
\begin{equation}
 \label{eq:F_semi}
 \beta \mathcal{F}_\mathrm{IV} \simeq \bigg( \frac{R_F^3}{D^2 L} \bigg)^\frac{1}{3\nu-1} \simeq \frac{D^2 L}{\xi_\mathrm{bulk}^3}
\end{equation}
\citep{GrosbergBook, Sakaue06, Cacciuto06, Jun07}, the above boundary condition is translated to $D = k \cdot \xi_\mathrm{bulk}$, which is in symmetry with the condition $L = k \cdot D$. Similarly, it is straightforward to show that the boundary condition $k=1$ between IV and V is identical to $y=x^{9/4}$ [Fig.~\ref{fig:phasediagram}(e)].

In regime $V$, the shape of the box resembles a thick slab. As the thickness of the slab reaches the size of the blob, {\i.e.,} $\xi_\mathrm{bulk} = L = k \cdot D$ [Fig.~\ref{fig:phasediagram}(e)], the two chains enter the dilute regime in a thin, closed slab [Fig.~\ref{fig:phasediagram}(g)], which is the dual regime of $II$. Thus, the transition between $V$ and $VI$ occurs when the chains start to feel the size of the slab, \i.e., at $D \sim R_\parallel \sim N^{1/2}\Phi^{-1/8}$, where the volume fraction is given by $\Phi \sim g/\xi_\mathrm{bulk}^3$, with $g \sim \xi_\mathrm{bulk}^{1/\nu}$. We thus obtain $y = x^6$ ($L = k \cdot D$).

\section{Self-consistent mapping between the Pincus and the close-packed chains.}
\label{appendix:pincus}
In the Pincus picture of a stretched chain, the chain breaks up into a series of blobs of size $\xi$ (Fig.~\ref{fig:renate}). The free energy cost can be estimated by counting the number of blobs,
\begin{equation}
 \label{eq:Pincus}
 \beta\mathcal{F}_\mathrm{P} \sim \frac{N}{g} \sim \frac{R}{\xi} \sim N\bigg( \frac{R}{N}\bigg)^\frac{1}{1-\nu},
\end{equation}

\noindent where $g \sim \xi^{1/\nu}$ is the number of monomers per blob and $R$ the end-to-end distance of the chain. 

Let us now imagine close-packing of the chain consisting of blobs of size $\xi$ in a cylinder of diameter $D$. If this process does not cost any additional free energy, using Eq.~\ref{eq:Pincus}, we can self-consistently obtain the confinement free energy accordingly as follows.
\begin{equation}
 \label{eq:cylinder}
 \beta\mathcal{F}_\mathrm{cyl} \sim \frac{L D^2}{\xi^3} \sim N\bigg( \frac{R}{N}\bigg)^\frac{1}{1-\nu} \sim \bigg( \frac{L D^2}{N^{3\nu}}\bigg)^\frac{1}{1-3\nu},
\end{equation}
where $L$ is the longitudinal size of the close-packed chain in the cylinder. Eq.~\ref{eq:cylinder} has the same scaling form of the free energy cost for confining a self-avoiding chain in a volume $V$,
\begin{equation}
 \label{eq:sphere}
 \beta\mathcal{F}_\mathrm{cyl} \sim \bigg( \frac{V}{R_g^3}\bigg)^\frac{1}{1-3\nu}
\end{equation}
(see Eq.~\ref{eq:F_semi}), and we thus have a self-consistent picture for interpreting compressed chain based on the Pincus chain. 

The essence of our analysis here is that the free energy for deformation costs $\sim k_BT/\mathrm{blob}$ regardless of the nature of deformation and, thus, is mappable one another (e.g., between stretching vs. confinement).

\bibliographystyle{apsrmplong}
\bibliography{entropy_reference_rmp}

\end{document}